\documentclass[aps,prl,twocolumn,superscriptaddress,showpacs]
{revtex4-1}

\usepackage{graphicx}% Include figure files
\usepackage{dcolumn}% Align table columns on decimal point
\usepackage{amsthm}
\usepackage{bm}% bold math
\usepackage{longtable}
\usepackage{color}
\usepackage{array}
\usepackage{booktabs}

\begin{document}

%\preprint{APS/123-QED}

\title{Observation of geometry dependent conductivity in two-dimensional electron systems}% Force line breaks with \\

\author{Dirk Backes}
\affiliation{Cavendish Laboratory, University of Cambridge, J. J. Thomson Avenue, Cambridge CB3 0HE, United Kingdom}
\author{Richard Hall}
\affiliation{Cavendish Laboratory, University of Cambridge, J. J. Thomson Avenue, Cambridge CB3 0HE, United Kingdom}
\author{Michael Pepper}
\affiliation{Department of Electronic and Electrical Engineering, University College London, Torrington Place, London WC1E 7JE, United Kingdom}
\author{Harvey Beere}
\affiliation{Cavendish Laboratory, University of Cambridge, J. J. Thomson Avenue, Cambridge CB3 0HE, United Kingdom}
\author{David Ritchie}
\affiliation{Cavendish Laboratory, University of Cambridge, J. J. Thomson Avenue, Cambridge CB3 0HE, United Kingdom}
\author{Vijay Narayan}
\affiliation{Cavendish Laboratory, University of Cambridge, J. J. Thomson Avenue, Cambridge CB3 0HE, United Kingdom}

\date{\today}% It is always \today, today,
\begin{abstract}

We report electrical conductivity $\sigma$ measurements on a range of two-dimensional electron gases (2DEGs) of varying linear extent. Intriguingly, at low temperatures ($T$) and low carrier density ($n_{\mathrm{s}}$) we find the behavior to be consistent with $\sigma \sim L^{\alpha}$, where $L$ is the length of the 2DEG along the direction of transport. Importantly, such scale-dependent behavior is precisely in accordance with the scaling hypothesis of localization~[Abrahams~\textit{et al.}, Phys. Rev. Lett. \textbf{42}, 673 (1979)] which dictates that in systems where the electronic wave function $\xi$ is localized, $\sigma$ is not a material-specific parameter, but depends on the system dimensions. From our data we are able to construct the "$\beta$-function" $\equiv (h/e^2) d \ln \sigma / d \ln L$ and show this to be strongly consistent with theoretically predicted limiting values. These results suggest, remarkably, that the electrons in the studied 2DEGs preserve phase coherence over lengths $\sim~10~\mu$m. This suggests the utility of the 2DEGs studied towards applications in quantum information as well as towards fundamental investigations into many-body localized phases.

\end{abstract}

\maketitle

\begin{figure*}
	\centering
	\includegraphics[width=6.5in]{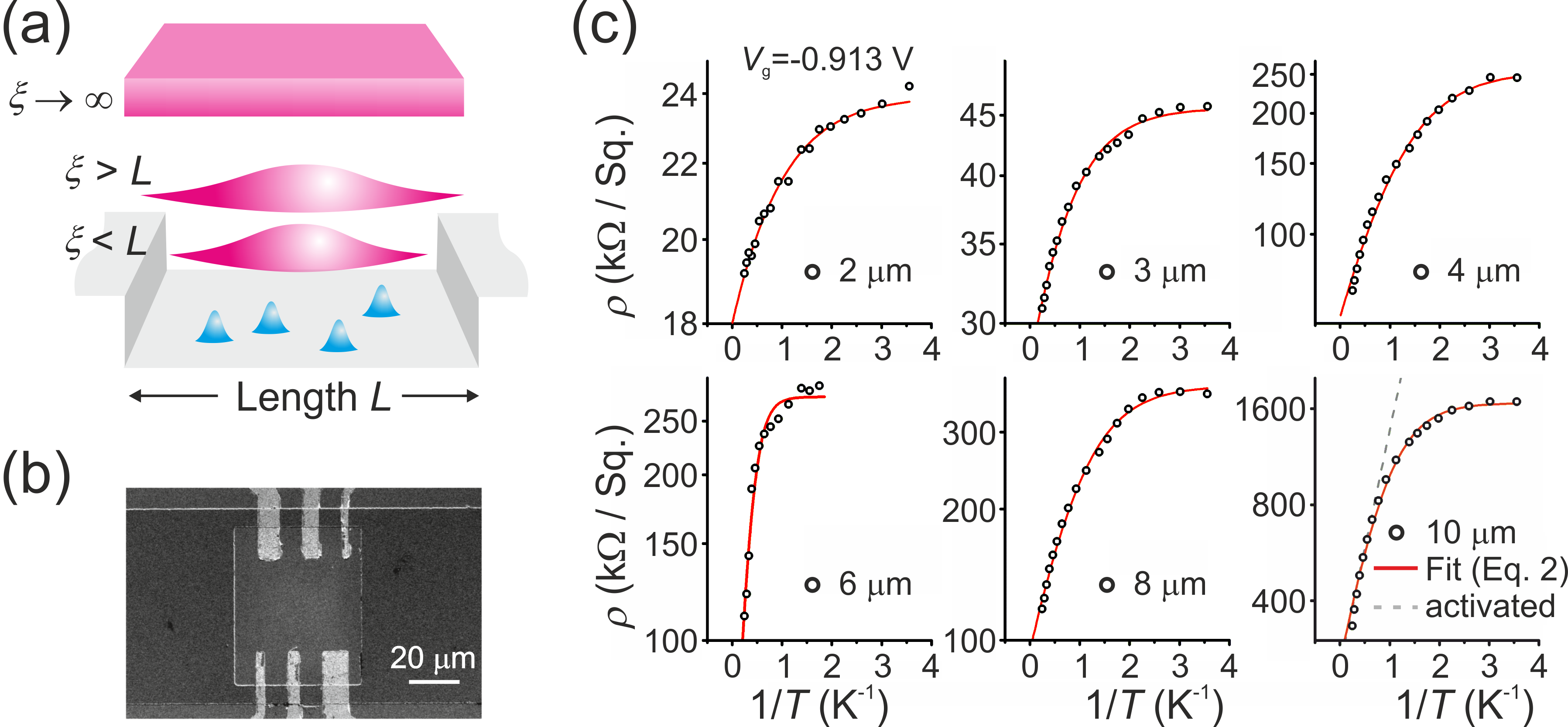}
	\caption{(a) If the conducting electrons in a material are extended, i.e., localization length $\xi \rightarrow \infty$, then the electrical conductance $G$ is finite and the conductivity $\sigma$ is well defined. However, for short-ranged states the relative extents of $\xi$ and the system $L$ decide the precise value of $G$. If $\xi > L$, then electrons can "bridge" the system and behave as though extended, i.e., induce a metallic character to the system. However, the metallicity is a finite-size effect, and in the large-$L$ limit, $G \rightarrow 0$. (b) Scanning electron microscope (SEM) image of the six top-gate-defined 2DEGs in a device similar to those used in this study. The light regions represent Ti-Au top-gate electrodes overlaid on a conducting mesa which is patterned into two parallel channels. Each top-gate can be individually addressed using a voltage source (see Ref.~\cite{SM} for details). (c) Each panel shows $\rho$ as a function of $1/T$ for a different sized 2DEG at an arbitrarily chosen gate voltage $V_{\mathrm{g}} = -0.913$~V. Here all the 2DEGs are ostensibly in the Anderson localized regime, and it is expected that $\rho \sim \exp(1/T)$ (shown as a broken line in bottom-right panel). However, the data seem consistent with the coexistence of metallic and insulating states. The solid lines are fits to Eq.~(\ref{two-resistor-model}).}
	\label{fig1}
\end{figure*}

The scaling hypothesis of localization~\cite{AALR}, formulated over thirty years ago, is a statement that the electrical conductivity $\sigma$ is lengthscale-dependent in finite systems where the conduction electrons are short-ranged or localized. This can be understood by considering electronic states with localization length $\xi$ in systems of different spatial extents: As depicted in Fig.~\ref{fig1}(a), if $\xi$ is greater than the linear extent of the system, then electrons are able to communicate across the system ends and there will be a finite conductance $G$ even at $T = 0$~K. However, this conductance will \textit{decrease} as the system  size increases, ultimately vanishing for infinitely large systems. On the other hand, if the electronic states are extended, $\xi \rightarrow \infty$, then even in the infinite system-size limit, $G \neq 0$. This intuitive picture is at the very heart of the scaling hypothesis which distinguishes between metallic and insulating states on the basis of the range of $\xi$: If the electronic states at the chemical potential $\mu$ are extended, then the system is a metal, but if they have a finite extent, the system is an insulator. In other words, the metallic state is defined by $\sigma$ independent of system dimensions, whereas the insulating state is characterized by $\sigma$ decaying with increasing system dimensions. This underlies the Anderson metal-to-insulator transition in which a "mobility-edge" in wave vector $k$-space demarcates short-ranged and long-ranged states~\cite{EversMirlinRMP2008}. 

However, since the scaling hypothesis was put forward, to our knowledge there have been \textit{no} experimental reports of length-dependent $\sigma$. In this paper, working with \textit{mesoscopic} GaAs-based 2DEGs of varying linear extent $L$, we provide the experimental demonstration of $\sigma$-scaling consistent with the scaling hypothesis. We continuously tune $\xi$ in the 2DEGs by applying a top-gate voltage $V_{\mathrm{G}}$ and observe a crossover from a regime in which the electrical resistivity $\rho \equiv 1/\sigma$ is independent of $L$ to one where it is strongly dependent on $L$. We find our results to be strongly consistent with the scaling predictions~\cite{AALR}.

In low-disorder two-dimensional (2D) systems $\xi \sim \ell\exp(k_{\mathrm{F}}\ell)$, where $\ell$ is the electronic mean free path and $k_{\mathrm{F}}$ is the Fermi wave vector. Using a 2DEG equipped with a top-gate electrode allows one to tune the carrier density $n_{\mathrm{s}}$ and thereby $k_{\mathrm{F}} = \sqrt{2 \pi n_{\mathrm{s}}}$. Furthermore, since $n_{\mathrm{s}}$ governs the degree to which any charged scattering centers are screened, this process also serves to vary $\ell$ which, in turn, can be estimated from the measured $\sigma$~\cite{SM}. Clearly, when $k_{\mathrm{F}}\ell >> 1$, $\xi$ can be macroscopically large, and this results in what is known as the "weakly localized" (WL) phase of electrons. The WL phase displays many outwardly metallic characteristics~\cite{AbrahamsRMP2001, Simmons_etal_PRL2000, Hamilton_etal_PRL2001, Uren_etal_JPhysC1981} including $d\sigma/dT \leq 0$ to the lowest achievable $T$~\cite{Kravchenko_PRB1995, LiPRL2003, PudalovPRL2003}, the hallmark of metallic conduction. When $k_{\mathrm{F}}\ell \approx 1$ experiments observe an abrupt crossover to the `strongly' or `Anderson' localized (AL) phase in which $\xi \sim a_{\mathrm{B}}^{\star}$, the effective Bohr radius in GaAs-based 2DEGs $\approx 11$~nm. In this regime $\sigma$ is completely suppressed, although at finite $T$ conduction occurs through phonon-assisted `hops'. This gives rise to $\sigma(T)\sim \exp(-(\Delta/k_{\mathrm{B}}T)^p)$, where $\Delta$ is the hopping energy, $k_{\mathrm{B}}$ is the Boltzmann constant and $p =$~1, 1/2 or 1/3 depending on whether the hopping is nearest-neighbour hopping~\cite{MottDavis}, hopping in the presence of the Coulomb gap~\cite{ES_JPhysC1975}, or variable-range hopping~\cite{MottDavis}, respectively. In other words, the sign of $d\sigma/dT$ can serve as a diagnostic to distinguish between metallic and insulating states. However, as we will directly show in this paper, the $T$-dependence alone is an insufficient test of metallicity. This is because, even in situations where $\xi \neq \infty$, (i.e., the system is, by definition, an insulator) $d\sigma/dT$ can be negative if $L \lesssim \xi$.

Experiments so far are consistent with the two limiting instances of $\xi >> L$ (WL) and $\xi << L$ (AL), neither of which, importantly, are expected to show $\sigma$-scaling. This is obvious in the AL or "hopping" regime since phonons, which mediate the hopping transport, exist homogeneously in space. The reasons for the absence of scaling behavior in the WL regime are, however, more subtle and perhaps linked to the macroscopic samples employed. Localization arises due to interference of the electronic wave function and thus relies crucially on phase coherence. The phase coherence length $\ell_{\mathrm{\phi}}$ is defined as the length over which the phase of the electron is completely randomized through inelastic interactions. Therefore scaling behavior is only expected when $L < \xi < \ell_{\mathrm{\phi}}$, a condition which may not have been rigorously met in earlier experiments~\cite{DaviesPepperKavehJPhysC1983}.

Here we perform a systematic size-dependence study of 2DEGs with varying $L$ and width $W$. As shown in Fig.~\ref{fig1}(b), our devices each contain six top-gate-defined 2DEGs with constant width $W$ and length $L$ ranging from 2 to 10~$\mu$m. We have fabricated devices with $W~=~$3~$\mu$m (D3), 9~$\mu$m (D9), and 11~$\mu$m (D11), and here we focus on the results from D9 and D11. Please refer to the Supplemental Material~\cite{SM} for details of the wafers used, device fabrication and measurement setup.

Figure~\ref{fig1}(c) shows resistivity $\rho \equiv 1/\sigma$ against $1/T$ for the six 2DEGs in D9 at gate voltage $V_{\mathrm{g}} =-0.913$~V. Here $\rho$ is evaluated as $R \times W/L$, where $R$ is measured in a quasi-four-terminal setup~\cite{SM}. The corresponding $\rho$ values are all $\gtrsim h/e^2$ and $k_{\mathrm{F}}\ell \lesssim 1$~\cite{SM} with $\ell$ determined from Drude theory. This would normally be the strongly localized regime where the $T$ dependence for an insulator can be expected. Remarkably, $d\sigma/dT\gtrsim 0$ for $T\lesssim 1$K, indicating the presence of metallic states as defined above~\cite{GhoshPRB2004, BaenningerPRL, KoushikPRB, NarayanPRB, NarayanJLTP}. While the device geometry has very little influence on the value of $T$ at which metallic conduction sets in, it is noteworthy that the $\rho(T \rightarrow 0)$ value is strongly device-dependent. This is despite the fact that the data are at the same value of $V_{\mathrm{g}}$, that the 2DEGs are located close to each other on the host wafer, and that they are all cooled down simultaneously under the same conditions. Importantly, while this behavior stands in stark contrast to the commonly observed 2D ``metal-to-insulator" transition, it suggests that the insulating and metallic states might be intimately linked.

\begingroup

\begin{table}
\label{tab1}
\begin{tabular}{|c|c|c|c|}
\hline
 $W \times L (\mu \mathrm{m^2})$& $\Delta/k_\mathrm{B}$ (K) &  $\rho_\mathrm{0} (k\Omega)$ & $\rho_\mathrm{L} (k\Omega)$ \\
\hline
9x10  & 2.25$\pm$0.08 & 273.87$\pm$14.41 & 1713$\pm$18.94 \\
9x8 & 1.68$\pm$0.06 & 114.22$\pm$4.14 & 409.51$\pm$4.93 \\
9x6 & 6.49$\pm$0.39 & 43.79$\pm$4.66 & 281.22$\pm$5.20\\
9x4 & 1.50$\pm$0.04 & 78.00$\pm$2.26 & 260.92$\pm$3.04 \\
9x3 & 1.64$\pm$0.06 & 63.21$\pm$1.37 & 45.14$\pm$0.22 \\
9x2 & 0.97$\pm$0.04 & 74.09$\pm$1.12 & 24.76$\pm$0.12 \\ 
\hline
\end{tabular}
\caption{Summary of fitting parameters for Fig.~\ref{fig1}c using Eq. \ref{ScalingHypothesis}.}
\end{table}
\endgroup

\begin{figure}
	\centering
	\includegraphics[width=3.25in]{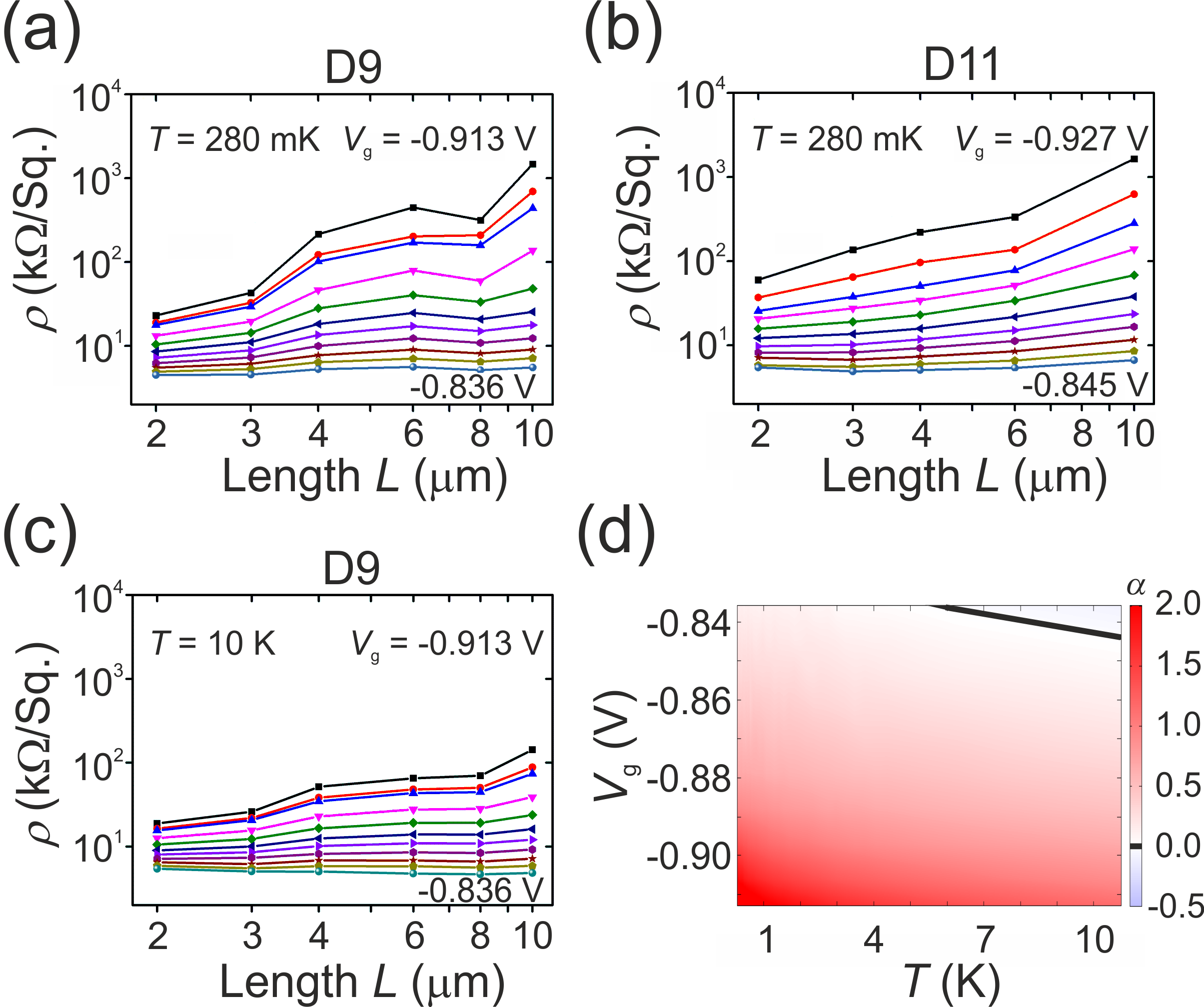}
	\caption{Geometry-dependent electrical characteristics. In (a) and (b) we see that $\rho \sim L^{\alpha}$ for both D9 and D11. (c) The power-law dependence of $\rho$ on $L$ persists upon heating to $T=10\,K$, but $\alpha$ is markedly reduced when compared to (a). (d) Phase diagram of $\alpha(V_{\mathrm{g}},T)$ for D9. The black line shows the locus of points where $\alpha$ becomes zero.}
	\label{fig2}
\end{figure}

Figures~\ref{fig2}(a) and \ref{fig2}(b) show $\rho$ as a function of $L$ at $T=0.28$\,K for D9 and D11, respectively. Interestingly, we find that the dependence seems broadly consistent with $\rho(L) \propto L^{\alpha}$. The exponent $\alpha$ decreases as $T$ increases and, as shown in Fig.~\ref{fig2}d, goes to zero at a $V_{\mathrm{g}}$-dependent $T$. We emphasize that the data shows the resistivity and not resistance, and thus any geometry-dependent characteristics are very unexpected. Figure~\ref{fig2}(c) shows $\rho(L)$ at 10\,K, and remarkably, we still find a clear and systematic $L$ dependence in $\rho$. It even appears that at the lowest $|V_{\mathrm{g}}|$ there is a slight negative slope [also seen in the top-right corner in Fig.~\ref{fig2}(d)], but this is a measurement artifact that becomes important only at low $|V_{\mathrm{g}}|$ (see Ref.~\cite{SM}). 

In all our measurements, we came across only one 2DEG (D11, $L$~=~8~$\mu$m) in which $\log\rho$ deviated by more than 10\% from $\alpha \log L$.

In the following, we analyze our experimental findings in light of the scaling hypothesis which states that

\begin{equation}
\label{ScalingHypothesis}
\sigma(L) = \frac{e^2}{2\pi \hbar}(k_{\mathrm{F}}\ell_{\mathrm{\circ}}) - \frac{e^2}{\hbar \pi^2}\ln(L/\ell_{\mathrm{\circ}})
\end{equation}

\begin{figure}
	\centering
	\includegraphics[width=3.25in]{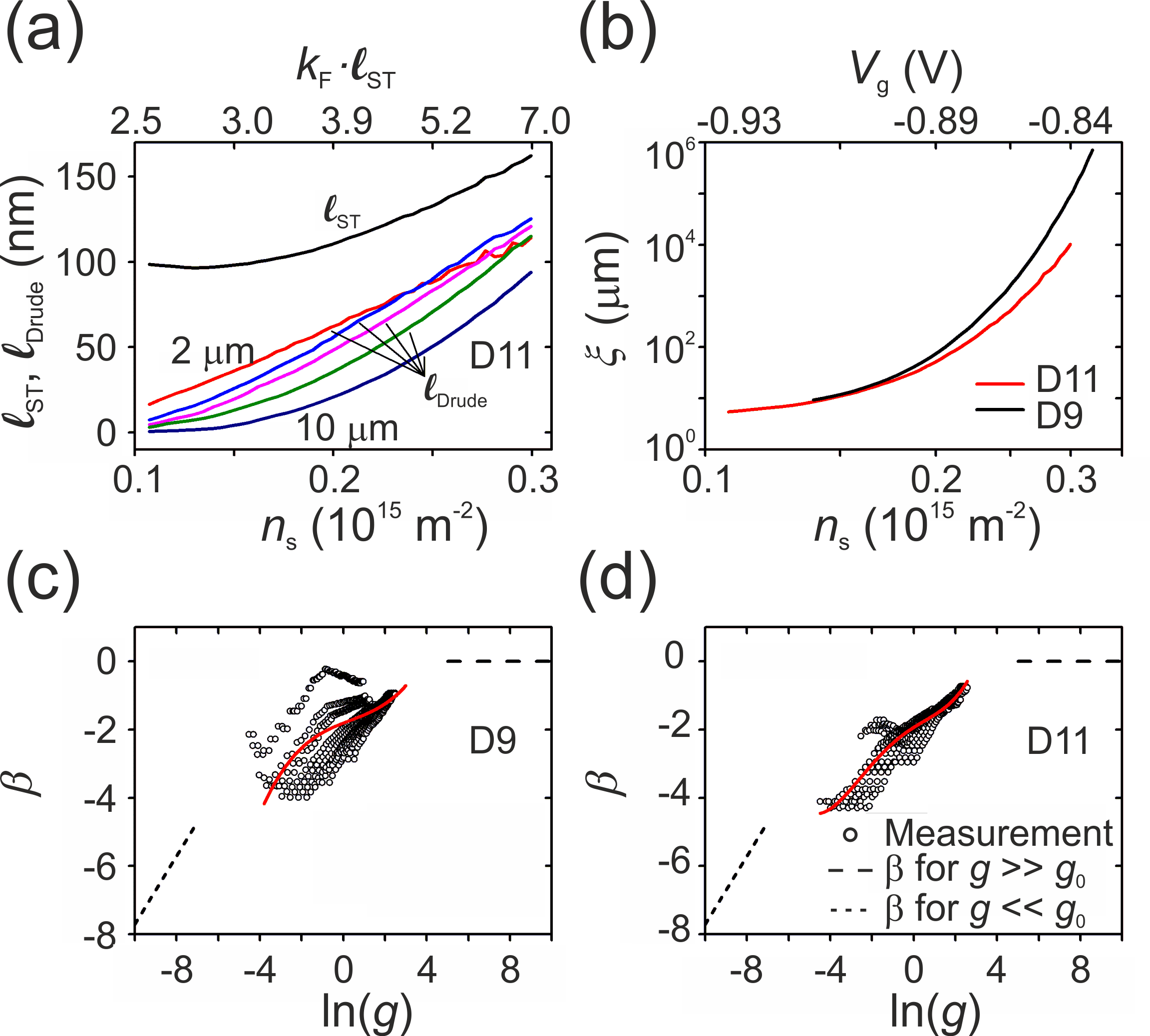}
	\caption{Localization and conductivity scaling in the 2DEGs. (a) The five lower curves show $\ell$ extracted using the Drude expression for $\sigma$~\cite{SM}, and the top curve shows $\ell$ estimated using the scaling hypothesis~\cite{AALR}. The corresponding $k_{\mathrm{F}}\ell$ values are shown along the top axis and lie well in the `metallic' regime. (b) Even at the lowest $n_{\mathrm{s}}$ where $\sigma \approx 0.01 e^2/h$ for $L=10 \mu$m, we find $\xi$ to be comparable to $L$. (c) and (d) show the scaling function $\beta$ as a function of $\ln g$, where $g\equiv \sigma/(e^2/h)$. The experimental data is in clear agreement with the theoretically predicted limiting values (shown as broken lines) for $g << g_{\mathrm{0}}$ and $g >> g_{\mathrm{0}}$, where $g_{\mathrm{0}} = e^2/h$. The red lines are guide to the eyes.}
	\label{fig3}
\end{figure}

Here the first term on the right is the conductivity at a microscopic length scale $\ell_{\mathrm{\circ}}$, and the second term is the size-dependent reduction in $\sigma$ arising due to the exponentially decaying envelope of $\xi$. The microscopic length scale is the smaller of $\ell$ and $\ell_{\mathrm{\phi}}$. We assume that $T$ is sufficiently low such that $\ell_{\mathrm{\phi}} > L, \ell$, an assumption which we will reexamine later. Therefore the first term is identically equal to the Drude conductivity $\sigma_{\mathrm{D}} = n_{\mathrm{s}}e^2\tau/m$~\cite{SM}, where $e$ is the electron charge, $\tau$ is the momentum scattering time, and $m$ is the electron effective mass in GaAs~=~0.067$m_{\mathrm{e}}$, with $m_{\mathrm{e}}$ being the bare electron mass. It is therefore important to note that $\sigma = \sigma_{\mathrm{D}}$ \textit{only} when $L/\ell$~=~1 and is suppressed for larger $L$. The length scale over which $\sigma \rightarrow 0$ is $\approx \ell \exp(\pi k_{\mathrm{F}}\ell/2)$, and this provides an estimate for $\xi$. On intermediate length scales, Eq.~(\ref{ScalingHypothesis}) clearly indicates that (incorrectly) identifying $\sigma(L)$ as $\sigma_{\mathrm{D}}$ results in an underestimate for $\ell$ and, importantly, that $(h/e^2)/\rho \neq k_{\mathrm{F}}\ell$. Indeed, as shown in Fig.~\ref{fig3}(a), upon fitting the measured $\rho(L)$ to Eq.~(\ref{ScalingHypothesis}), we obtain $\ell \approx$~100~nm at the lowest accessible $V_{\mathrm{g}}$, which is significantly greater than the nominally obtained $\ell$ values. We are able to map from $V_{\mathrm{g}}$ to $n_{\mathrm{s}}$ by tuning the device to the quantum Hall regime and observing edge-state reflections as $V_{\mathrm{g}}$ is decreased~\cite{BaenningerPRB2005} and thereby ascertain that the corresponding $n_{\mathrm{s}}$~=~1.4~$\times$~10$^{14}$~m$^{-2}$. This results in $k_{\mathrm{F}}\ell = \sqrt{2\pi n_{\mathrm{s}}}\ell_{\mathrm{ST}} \approx$~3, even though the measured $\rho$ is orders of magnitude greater than $h/e^2$ and $k_{F}\ell_{\mathrm{Drude}}< 1$ (see Fig.~\ref{fig2}). For $k_{\mathrm{F}}\ell = 3$, we estimate $\xi \approx$~11~$\mu$m which, crucially, is comparable to the device dimensions. Similar results are obtained for D11, and $\xi$ is plotted as a function of $n_{\mathrm{s}}$ in Fig.~\ref{fig3}(b). 

Figs.~\ref{fig3}(c) and \ref{fig3}(d) provide a complementary look at the scaling behavior in our data by examining the scaling function $\beta \equiv d\ln g/d\ln L$ as a function of $\ln g$, where $g \equiv \sigma/(e^2/h)$. $\beta$ is evaluated from each pair of neighboring points in Figs.~\ref{fig2}(a) and \ref{fig2}(b). The general trend in $\beta$ (solid red line) agrees well with the theoretical limits of $\beta$ for very large and small $\ln g$. These theoretical limits arise from a combination of dimensional considerations in the low-disorder ($g\gg 1$) regime and some basic assumptions about the overlap of localized states in the high-disorder ($g\ll 1$) regime.  In the former, where disorder and scattering are weak, the electronic wave function will have a very large extent, and it is reasonable to expect that $\sigma(L)$ is intensive $\sim GL^{d-2}$, where d is the dimensionality of the system under study. In the latter, where disorder is strong and the electronic wave function is localized, conduction is governed by the spatial overlap of neighboring states. However, such localized states cannot cumulatively result in an extended state since states in close spatial proximity are necessarily widely separated in energy. Thus $\sigma(L)$ is exponentially suppressed $\sim \exp (−L/\xi)$, independent of dimensionality. These expressions for $\sigma$ provide the theoretically expected limits in $\beta$, which in 2D reduce to $\beta=0$ for $g\gg1$ and $\beta=\ln g$ for $g\ll1$. We find the averaged $\beta$, obtained from our measurements, exactly in the range in between the theoretically expected limiting values. It is noteworthy that Fig.~\ref{fig3} provides evidence of finite $\xi$ within the WL regime where $3 < k_{\mathrm{F}}\ell_{\mathrm{ST}} < 7$ where the subscript denotes "scaling theory".

Thus the picture emerges that the 2DEGs studied are, in fact, in the weakly localized regime but with $\sigma$ significantly reduced due to the finite extent of $\xi$. Therefore, the weak dependence and even positive slope of $\rho$ against $T$ are entirely expected. The question then arises as to why above 1~K the 2DEGs show activated transport. The point here is that the metallic character below 1~K is imparted by the relatively long $\xi \gtrsim L$ electronic states at $E = \mu$, but states with $E << \mu$, which nominally do not contribute to transport due to phase space restrictions, are continually hopping due to inelastic interactions with phonons. These therefore provide an additional transport channel with an activated form. We thus propose a simple "parallel-resistor" model to understand the $T$-dependence of $\rho$ in which the conducting states (at $E = \mu$) and hopping states (at $E < \mu - k_{\mathrm{B}}T$) conduct in parallel:

\begin{equation}
\label{two-resistor-model}
\frac{1}{\rho} = \frac{1}{\rho_{\mathrm{L}}} + \frac{1}{\rho_{\mathrm{0}}}\exp{(-\Delta/k_{\mathrm{B}}T)}.
\end{equation}

Here $\rho_{\mathrm{L}}$ is the contribution due to the effectively extended states assumed to be $T$ independent, and the second term on the right is the hopping term. As shown in Fig.~\ref{fig1}(c), we are able to obtain excellent fits to the data using $\rho_{\mathrm{L}}$, $\rho_{\mathrm{0}}$ and $\Delta$ as fitting parameters. Values of the resulting fit parameters are listed in Table~\ref{tab1}. A noteworthy, though small, feature that the model does not capture is the mildly positive $d\rho/dT$ seen at $1/T >$~2~K$^{-1}$ in the lower-middle panel of Fig.~\ref{fig1}(c) (see also Ref.~\cite{BaenningerPRL}). However, this is trivially so due to the assumption of constant $\rho_{\mathrm{L}}$ which disregards effects such as electron screening~\cite{DasSarmaHwang_PRL1999} and interelectron interactions~\cite{Zala_etal_PRB2001}, including which will, no doubt, result in more accurate models.

We now return to our assumption that $\ell_{\mathrm{\phi}} > L$, which is a necessary requirement for coherent electron interference and, thus, for localization effects to manifest. As $T \rightarrow \infty$ phase coherence is lost, and it is to be expected that localization phenomena be suppressed. We see clear evidence of this in Fig.~\ref{fig2}(c), where $\alpha$ gradually diminishes to 0 as $T$ increases. 

In addition, it is observed that at a fixed $T$, decreasing $|V_{\mathrm{g}}|$ also diminishes $\alpha$, which is consistent with an Anderson-like transition to a metallic state~\cite{Basko_etal, NandkishorePRB2014}. However, the $V_{\mathrm{g}}$-dependence can be seen simply as a consequence of $\xi >> L$. In addition, the thermopower $S$ of similar 2DEGs displays strong oscillations and even sign changes~\cite{NarayanNJP} which might have their origin in phase coherent transport~\cite{LesovikKhmelnitskii}. Thus, there are several indications that electrons retain phase coherence over the length of the devices studied. This is a remarkable observation given that (i) the largest 2DEGs have $L\sim 10~\mu$m, which is significantly longer than conventionally measured $\ell_{\mathrm{\phi}}$ (see, for example, Ref.~\cite{Ferrier_etal_PRL2004}), and (ii) the $L$ dependence is seen even at 10 K. We comment on why this might be so further on in the paper but at this stage emphasize the strong applicability of the systems studied in quantum information schemes.

Before presenting our concluding remarks, we first consider the important issues of (i) the background disorder potential the 2DEGs reside in~\cite{TripathiKennetPRB2006, TripathiKennetPRB2007, NeilsonHamitlonPRB2010} and (ii) interelectron interactions. Implicit in our analysis based on the scaling hypothesis is the assumption that the background disorder experienced by the various 2DEGs is statistically homogeneous. However, we believe this assumption to be amply justified by the systematic $\rho$ vs $L$ trend observed in all three sets of 2DEGs: D9, D11 and D3~\cite{SM}. We have also found this trend to be reproducible between cooldowns, albeit with marginally different pinch-off characteristics~\cite{SM}. Thus, it seems reasonable to believe that the statistical degree of inhomogeneity in the disorder is small, perhaps responsible for the departures from perfect linearity in Figs.~\ref{fig2}(a) and \ref{fig2}(b). It is also conceivable that "fluctuations" in $\xi$ due to the mesoscopic nature of the 2DEGs are influencing transport~\cite{SchreiberPRL1991, SoukoulisEconomouWRM1999, MirlinPhysRep2000, CuevasKravtsovPRB2007}. In other words, it is possible that the small scale of the 2DEG facilitates observation of certain ballistic electron trajectories which are not observable at longer length scales. However, we believe we have minimized the effect of such nonergodicities by (i) working at $V_{\mathrm{g}}$ values where $\ell < L$ [see Fig.~\ref{fig3}(a) and also Ref.~\cite{SM}] and (ii) averaging our data over long times before recording~\cite{SM}. The second important point to consider is that of interelectron interactions which, importantly, must be present in the 2DEGs since the interaction parameter $r_{\mathrm{s}}$, which is the ratio of the Coulomb energy $E_{\mathrm{C}}$ and kinetic energy $E_{\mathrm{K}}$ of the electrons =~$1/(a_{\mathrm{B}}^{\star}\sqrt{\pi n_{\mathrm{s}}})$, attains values as large as 5 in our studies. However, as was demonstrated recently in Ref.~\cite{Mard_etal_PRL2015}, qualitative changes in $\beta$ are not expected even in the presence of strong interactions, and this is consistent with our findings. It will be interesting to understand whether the recently observed strong enhancement in the magnitude of $S$ measured in similar 2DEGs~\cite{NarayanPRB} or the observed violation of the Mott formula~\cite{NarayanPRB, NarayanNJP}, where $S$ and $\rho$ were observed to oscillate asynchronously, reflect strong interaction effects or not. Lastly, we also wish to point out the similarities between our experimental results and the phenomenology of many-body delocalized phases in translationally-invariant 2D systems~\cite{NandkishorePRB2014}. While it is debatable whether our experimental system stringently fulfills the criteria for many-body delocalization, namely complete isolation from the environment, we note that this is certainly consistent with the lack of electron decoherence even at $\approx~10~$K.

In conclusion we emphasize that the observed $L$ dependence of $\sigma$ in mesoscopic 2DEGs is strongly consistent with the scaling hypothesis, which in turn suggests that the 2DEGs are, in fact, in the $k_{\mathrm{F}}\ell >$~1 regime, but perceptibly Anderson localized, i.e., with $\sigma$ suppressed due to the finite $\xi$. In macroscopic 2DEGs where $L \gg \ell_{\mathrm{\phi}}$, conductivity scaling may not be apparent since blocks of size $\ell_{\mathrm{\phi}} \times \ell_{\mathrm{\phi}}$ contribute in an incoherent fashion. Nevertheless, it is important to note that as long as $\ell_{\mathrm{\phi}} > \ell$ the conductivity of such a block \textit{must} be diminished from its value at size $\ell \times \ell$, rendering imprecise the identification that $k_{\mathrm{F}}\ell = (h/e^2)/\rho$. The observation of scaling-like behavior at length scales of several $\mu$m and at temperatures of $\approx$~10\,K suggest the system under study to be remarkably robust to decoherence effects. While we do not fully understand why this might be, we speculate that this has to do with the specific device geometry~\cite{SM} in which the ohmic contacts are at a large spatial separation from the 2DEGs being studied. Thus, the primary link the 2DEGs have to the environment is the tenuous low-$T$ electron-phonon coupling which might be further weakened due to the narrow mesa widths employed~\cite{Banerjee:2015, Backes:2016}. Importantly, this opens up several possibilities towards studying many-body localized electron phases~\cite{Basko_etal}.

We acknowledge funding from the Leverhulme Trust, UK and the Engineering and Physical Sciences Research Council (EPSRC), UK. We also acknowledge D. Joshi for assistance with device fabrication. DB and VN acknowledge useful discussions with Margarita Tsaousidou, Chris Ford, Charles Smith, Moshe Kaveh and Richard Berkovits. Supporting data for this paper is available at the DSpace@Cambridge data repository (https://www.repository.cam.ac.uk/handle/ 1810/252722).


\setcounter{figure}{0} \renewcommand{\figurename}{Fig. S}

\onecolumngrid

\vspace{12pt}

\begin{center}

\large \textbf{Observation of geometry dependent conductivity in two-dimensional electron systems \\ Supplementary Material}

\end{center}

\twocolumngrid

\section{Equivalence of $\sigma_{\mathrm{D}}/(e^2/h)$ and $k_\mathrm{F}\ell$}

The Drude conductivity $\sigma_{\mathrm{D}} = n_{\mathrm{s}}e^2\tau/m$, where the symbols are defined in the main text. The momentum relaxation time $\tau = v_{\mathrm{F}}\ell$, where $v_{\mathrm{F}}$ is the Fermi velocity $= \hbar k_{\mathrm{F}}/m$. Substituting the expression for $v_{\mathrm{F}}$ into $\sigma_{\mathrm{D}}$ and rearranging, we arrive at the expression: $\sigma_{\mathrm{D}}/(e^2/h) = k_{\mathrm{F}}\ell$.

However, the point we make is that according to the scaling hypothesis (Eq.~\,(1) in the main text) the experimentally measured $\sigma$ is almost always smaller than $\sigma_D$ due to the finite extent of $\xi$. Therefore, naively estimating $\ell$ as $\sigma /(e^2/h k_{\mathrm{F}})$ would result in an underestimate for $\ell$, and the appropriate manner in which to estimate $\ell$ is from fitting the size dependence of $\sigma$ to Eq.\,(1) of the main text.

\section{Experimental methods}

\subsection{Wafer details and device fabrication}

The wafers used in this experiment are MBE-grown $\delta$-doped structures in which the 2DEG resides 300\,nm below the surface. The $\delta$-dopants lie 40\,nm above the 2DEG. At 4~K, the mobility of a macroscopic ($L\times W = 1000$\,$\mu$m\,$\times$\,100\,$\mu$m) Hall bar sample was measured to be 220\,m$^2$/Vs with carrier density $n_s = 2.1 \times 10^{15}$\,m$^{-2}$.

Devices were fabricated using three stages of optical lithography. First, the conducting mesa was defined using a wet chemical etch, after which Au-Ge-Ni Ohmic contacts were deposited by therml evaporation. These were annealed at 450$^{\circ}$  in an atmosphere of forming gas in order for electrical contact to the buried 2DEG. Finally, Ti-Au top-gates were thermally evaporated onto the patterned sample surface. As shown schematically in Fig.\,(S\,1), each device contains six top-gate-defined 2DEGs. The mesa was defined with two parallel arms, rather than one long one in which the 2DEGs would all be in series. This was done in order to keep the 2DEGs in close proximity to each other (to minimize variations in the background disorder), and also to avoid the large series resistance associated with long, narrow sections of mesa (see next section for further details).

\subsection{Electrical measurements}

\begin{figure}
	\centering
	\includegraphics[width=3.25in]{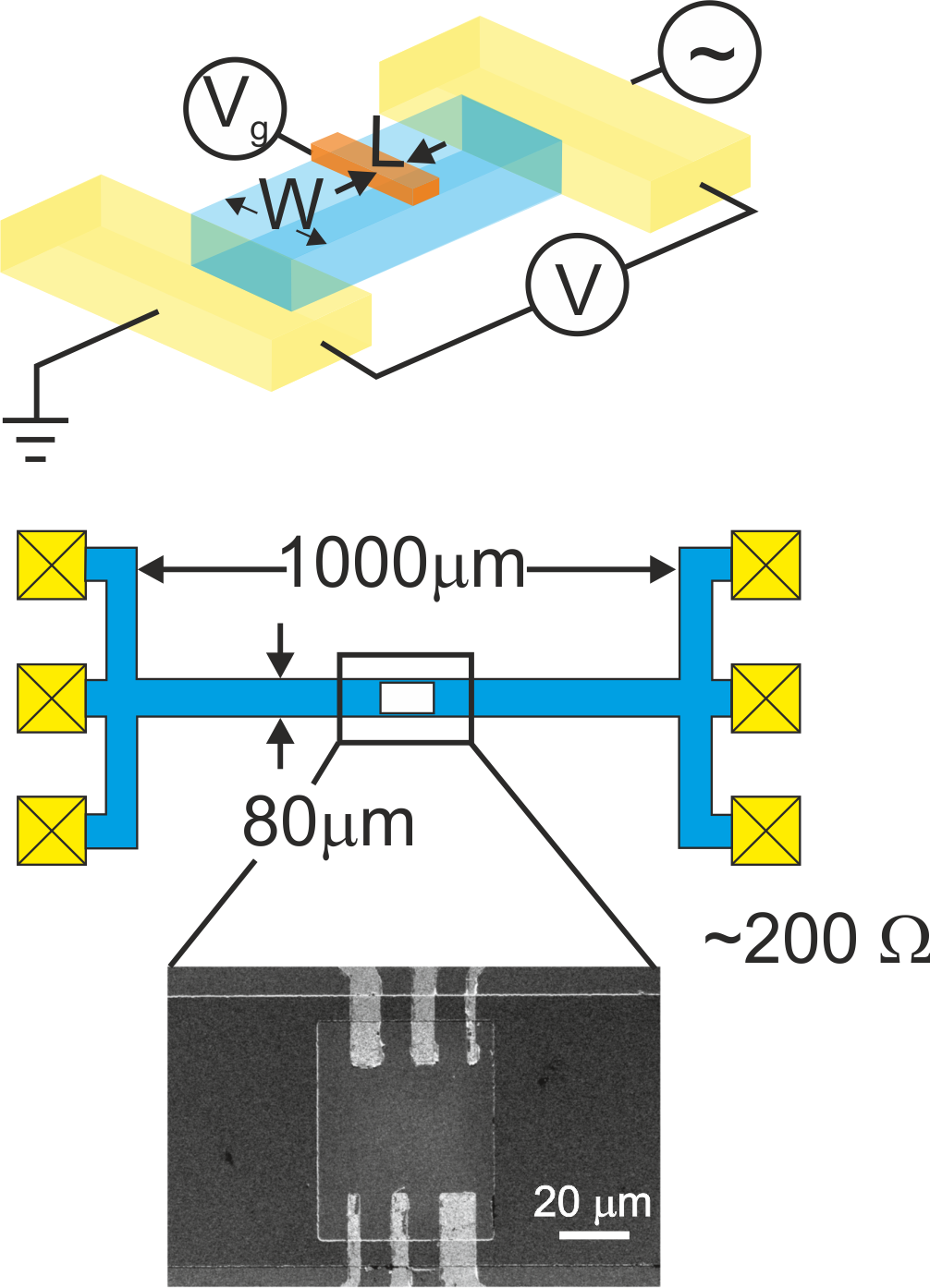}
	%\captionsetup{justification=raggedright,singlelinecheck=false}
	\caption{The top panel shows a schematic of the quasi-four terminal measurement setup. The blue regions indicate the conducting mesa, the yellow regions indicate ohmic contacts and the orange represents the top-gate. Regions outside the top-gated area contribute a constant $R \approx 200\,\Omega$ to the measurement. The middle panel shows a top-view of the device layout focusing on the central area containing the 2DEGs under study. The bottom panel is a scanning electron microscope (SEM) image of the six 2DEGs (light areas depict Ti-Au gates overlaid on the patterned mesa).}
	\label{figS1}
\end{figure}

The measurements were performed between 0.3\,K and 10\,K in a He-3 cryostat equipped with a superconducting magnet. Devices were cooled down from room temperature to 4\,K over a period of 20\,hours and only 2DEGs that were cooled down simultaneously were compared against each other. Each 2DEG was addressed individually with a DC voltage source. When a particular 2DEG was being measured, a large negative $V_{\mathrm{g}}$ (-5\,V) was applied to all three gates on the adjacent mesa arm in order to completely cut off any parallel conduction. We ascertained that the resistance of the adjacent arm was $> 10$\,G$\Omega$ using a Keithly 236 Source-Measure unit.

Electrical measurements were made in a \textit{quasi}-four-terminal setup (see below) using an excitation current $I =$~100\,pA at frequency $f = 7$\,Hz. We ascertained that there was no appreciable joule heating by increasing $I$ to 1\,nA, and noting no change in the experimental data. Our measurements are performed by sweeping the gate voltage $V_{\mathrm{g}}$ slowly such that each data point is averaged for several (10 -- 100) seconds before recording.

There are several factors that we had to carefully consider when measuring the mesoscopic 2DEGs:\\
1. From the device design it is clear that there are ungated sections of the mesa that contribute to the resistance measurement, i.e., there is an extra `lead' resistance $R_{\mathrm{L}}$. At $V_{\mathrm{g}} = 0$\,V, $R_{\mathrm{L}} \approx R_{\mathrm{m}}$ (the measured resistance), the approximation arising from not excluding the 2DEG area. However, as seen from Fig.~S~1, this corresponds to an error of $< 1\,\%$, corresponding to the length of the 2DEG (at most 10\,$\mu$m) divided by the length of the entire mesa (1000\,$\mu$m). This therefore allows us to subtract $R_L$ ($\approx$\,200\,$\Omega$ in all the devices) to estimate the true 2DEG resistance $R$. Clearly this approximation becomes less reliable as $|V_{\mathrm{g}}| \rightarrow 0$, due to which we restrict our analysis to large $|V_{\mathrm{g}}|$.\\
2. A second reason to restrict the analysis to high $|V_{\mathrm{g}}|$ is to minimize any ballistic electron effects that might be significant when $\ell \sim L$. At $V_{\mathrm{g}} = 0$\,V, the mobility and carrier density correspond to $\ell \approx 17$\,$\mu$m which is larger than the largest 2DEG investigated in this study. By confining the analysis to $|V_{\mathrm{g}}| \geq 0.84$\,V, the 2DEGs are always in a regime where $\ell < 0.1L$, i.e., where the electronic motion is diffusive.\\
3. And finally, we note that there will be electric-field fringing at the edges of the top-gate defined 2DEGs. However, we expect these to be of the order of the 2DEG setback distance =\,300\,nm and, moreover, that these will contribute a constant $R$ offset to each 2DEG and therefore, not influence the results in a major way.\\
The 2DEG resistivity $\rho$ is defined as $R \times W/L$ and based on the above arguments, we are confident that this is a meaningful definition bereft of any artifacts due to inhomogeneities, ballistic electron trajectorites, or even surface/boundary scattering. The last of these follows from point 2 where we argue that the electrons have a well-defined diffusivity.

%\begin{widetext}
\section{Extra supporting data}

%\begin{minipage}{\linewidth}
%\centering
%	\includegraphics[width=6.5in]{S2_D9.png}
%	\captionof{figure}{The figure shows data from device D9 on two separate cooldowns. The left panel shows Fig.~2a from the main text. The clear trend is unmistakable, but equally noteworthy is the fact that the pinch-off characteristics are different between cooldowns, sugggesting that the disorder profile depends crucially on the cooldown. The missing data points on the right panel correspond to devices that were not measured since the experimental run needed to be terminated prematurely due to technical issues.}
%	\label{figS2}
%\end{minipage}

\begin{figure*}
	\centering
	\includegraphics[width=6.5in]{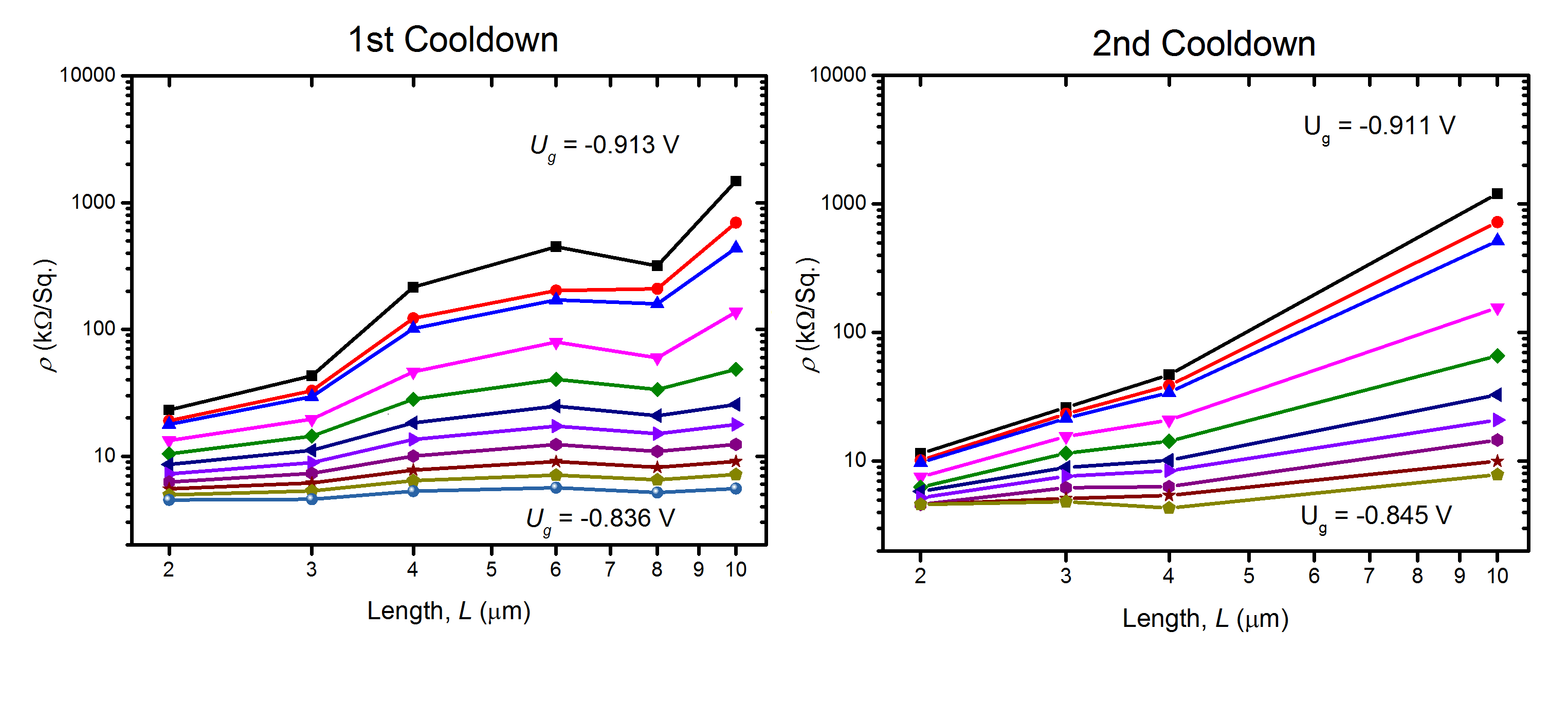}
	\caption{The figure shows data from device D9 on two separate cool downs. The left panel shows Fig.~2(a) from the main text. The clear trend is unmistakable, but equally noteworthy is the fact that the pinch-off characteristics are different between cool downs, suggesting that the disorder profile depends crucially on the cool down. The missing data points on the right panel correspond to devices that were not measured since the experimental run needed to be terminated prematurely due to technical issues.}
	\label{figS2}
\end{figure*}

%\begin{figure}
%\centering
%\begin{minipage}{.45\textwidth}
%  \centering
%  \includegraphics[width=3.25in]{S3_D3.png}
%  \captionof{figure}{The figure shows $\rho$ vs $L$ for device D3 in which the width of the mesa is 3~$\mu$m. This data is taken in a dilution refrigerator at 100~mK and shows the same trend observed in D9 and D11 where $\rho$ grows with increasing $L$.}
%  \end{minipage}%
%\begin{minipage}{.45\textwidth}
%  \centering
%  \includegraphics[width=3.0in]{S4_rho_vg_BField.png}
%  \captionof{figure}{The figure shows $\rho$ vs $V_{\mathrm{g}}$ for D11 in the presence of a perpendicular $B$-field = 0~T, 3~T, 6~T and 9~T. In addition to the strong positive magnetoresistance, one observes weak oscillations in $\rho$ which likely correspond to Shubnikov-de Haas oscillations that arise when the chemical potential is tuned through Landau levels.}
%  \end{minipage}
%\end{figure}

\begin{figure}
	\centering
	\includegraphics[width=3.25in]{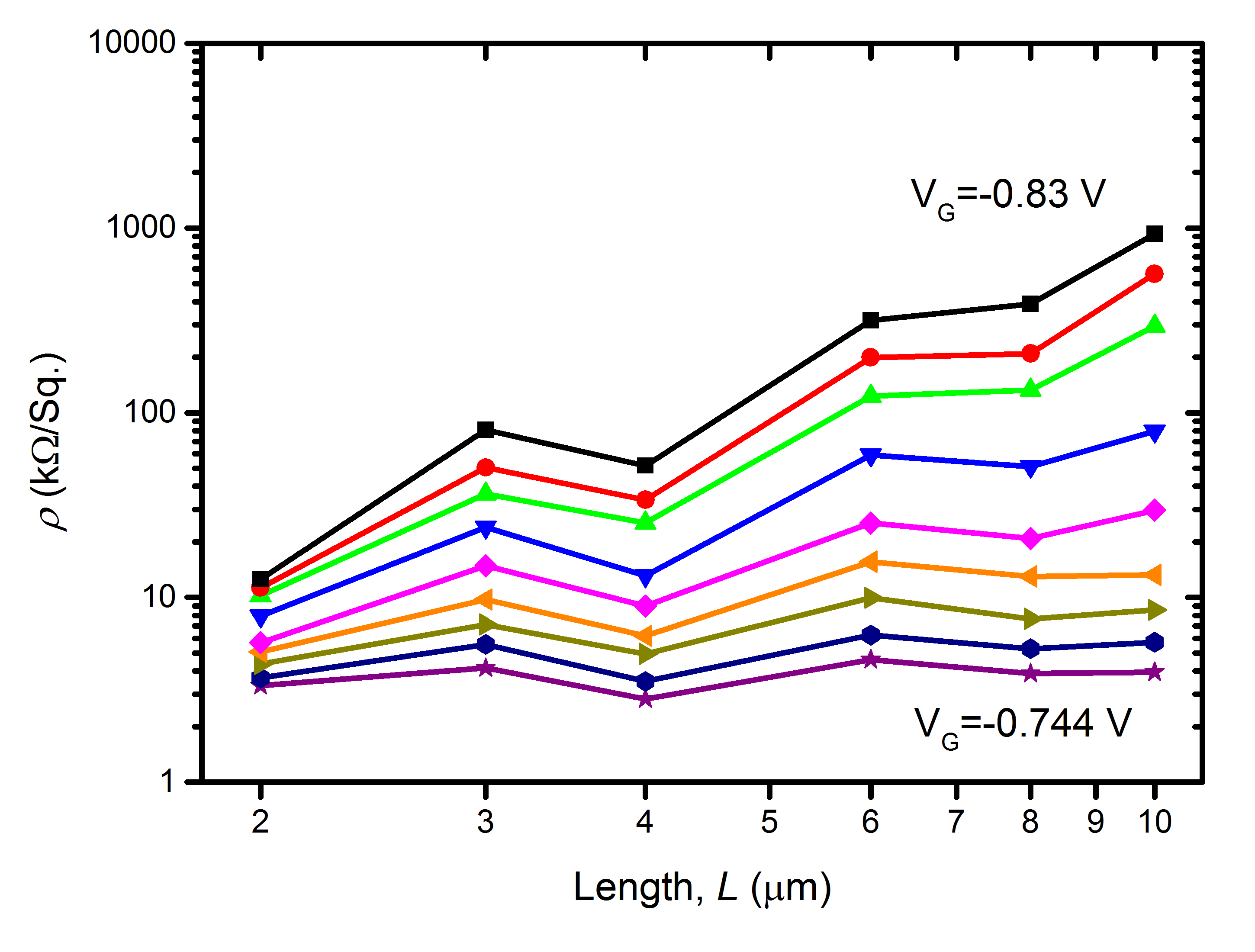}
	\caption{The figure shows $\rho$ vs $L$ for device D3 in which the width of the mesa is 3~$\mu$m. This data is taken in a dilution refrigerator at 100\,mK and shows the same trend observed in D9 and D11 where $\rho$ grows with increasing $L$.}
	\label{figS3}
\end{figure}

\begin{figure}
	\centering
	\includegraphics[width=3.25in]{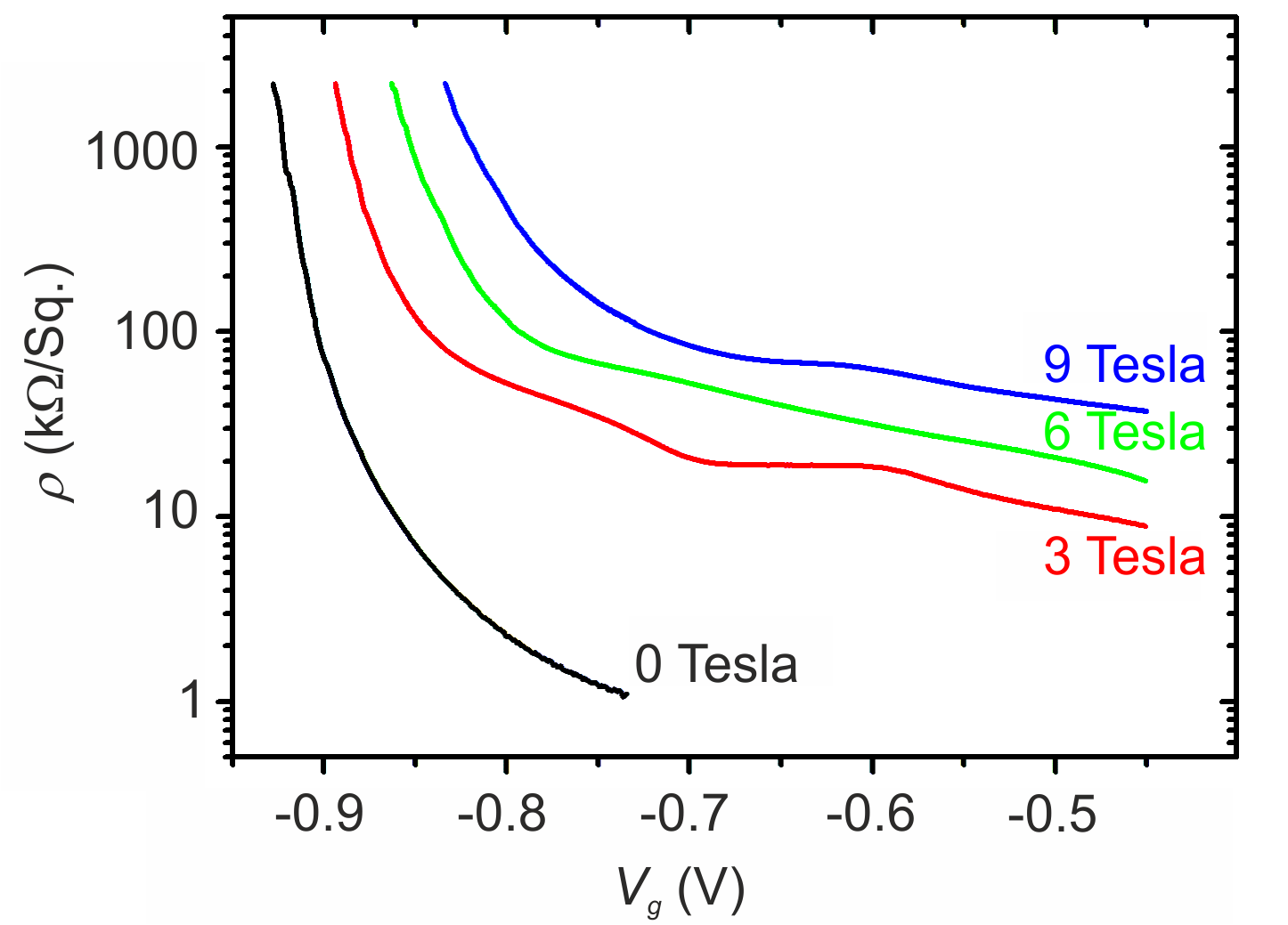}
	\caption{The figure shows $\rho$ vs $V_{\mathrm{g}}$ for D11 in the presence of a perpendicular $B$-field = 3\,T, 6\,T and 9\,T. In addition to the strong positive magnetoresistance, one observes weak oscillations in $\rho$ which likely correspond to Shubnikov-de Haas oscillations that arise when the chemical potential is tuned through Landau levels (see also Fig.~S\ref{figS5}).}
	\label{figS4}
\end{figure}

\begin{figure}
	\centering
	\includegraphics[width=3.25in]{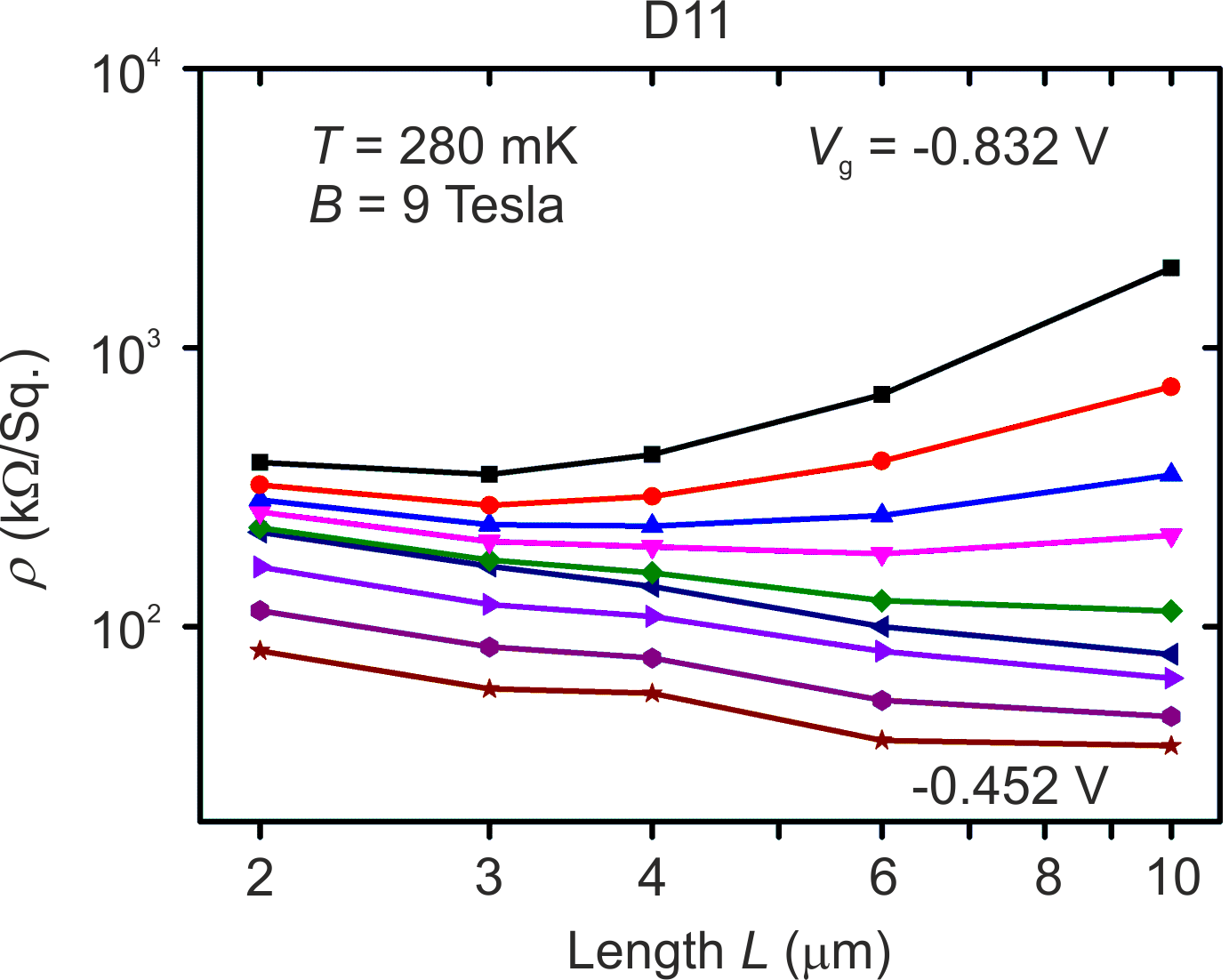}
	\caption{The figure shows $\rho$ vs $L$ for device D11 in which the width of the mesa is 11\,$\mu$m. This data is taken at 280\,mK and shows the the power-law dependence of $\rho$ on $L$ persists in the presence of a perpendicular magnetic field of 9 Tesla. The exponent $\alpha$ is seen to assume positive as well as negative values. The negative values are a consequence of Shubnikov-de Haas minima.}
	\label{figS5}
\end{figure}

\end{document}